\documentclass[]{aastex631}

\begin{document}

\title{A Potential Dynamical Origin of The Galactic Disk Warp: The Gaia-Sausage-Enceladus Major Merger}

\author[0000-0002-0592-7660]{Mingji Deng}
\affiliation{School of Astronomy and Space Sciences, University of Chinese Academy of Sciences, Beijing 100049, P.R. China}

\author[:0000-0002-3954-617X]{Cuihua Du}
\affiliation{School of Astronomy and Space Sciences, University of Chinese Academy of Sciences, Beijing 100049, P.R. China}

\author[0000-0001-7949-3407]{Yanbin Yang}
\affiliation{GEPI, Observatoire de Paris, Universite PSL, CNRS, Place Jules Janssen, 92195 Meudon, France}

\author[0000-0001-6762-5599]{Jiwei Liao}
\affiliation{School of Astronomy and Space Sciences, University of Chinese Academy of Sciences, Beijing 100049, P.R. China}

\author{Dashuang Ye}
\affiliation{School of Astronomy and Space Sciences, University of Chinese Academy of Sciences, Beijing 100049, P.R. China}

\correspondingauthor{Cuihua Du}

\email{ducuihua@ucas.ac.cn}



\begin{abstract}

 Previous studies have revealed that the Galactic warp is a long-lived, nonsteady, and asymmetric structure. There is a need for a model that accounts for the warp's long-term evolution. Given that this structure has persisted for over 5 Gyrs, its timeline may coincide with the completion of Gaia-Sausage-Enceladus (GSE) merger. Recent studies indicate that the GSE, the significant merger of our Galaxy, was likely a gas-rich merger and the large amount of gas introduced could have created a profound impact on the Galactic morphology. This study utilizes GIZMO simulation code to construct a gas-rich GSE merger. By reconstructing the observed characteristics of the GSE, we successfully reproduce the disk warp and capture nearly all of its documented features that aligns closely with observational data from both stellar and gas disks. This simulation demonstrates the possibility that the single major merger could generate the Galactic warp amplitude and precession. Furthermore, the analysis of the warp's long-term evolution may offer more clues into the formation history of the Milky Way.

\end{abstract}

\keywords{Galaxy structure (622); Milky Way disk (1050); Milky Way dark matter halo (1049); Galaxy mergers (608); Hydrodynamical simulations (767)}


\section{Introduction} \label{sec:intro}

   The most common asymmetrical structure in many disk galaxies is the disk warp, as documented in previous studies \citep{SS90, SS03, Reshetnikov}. The Milky Way, a typical example of disk and spiral galaxies, also exhibits a clear disk warp, confirmed by various works \citep{Kerr, Levine, Freudenreich, Chen}. Many mechanisms of the Galactic warp have been proposed, including: inflow of intergalactic matter into the halo \citep{OB, QB, JB}; inflow directly onto the Galactic disk \citep{LC}; magnetic fields that exist between galaxies \citep{BJ}; interaction between satellite galaxies like Sagittarius \citep{Bailin} or Magellanic Clouds \citep{WB} with the disk; and the bending instability and self-excited warps or internally driven warps of the Galactic disk \citep{RP, SD}. However, none of these hypotheses have been quantitatively confirmed as the definitive origin of the Milky Way's warp. Recent studies have suggested that the MW's S-shaped warp may be a long-standing structure, existing for over 5 Gyrs \citep{Roskar, LC14, Li23}. Constructing a model of long-term evolution could provide insights into the origin of the Galactic warp, and such a period may coincide with the time that Gaia-Sausage-Enceladus (GSE) merger completed.

   Lines of evidence have shown that the GSE, which is considered as the last major merger of our Galaxy, constitutes the bulk of the inner halo \citep[e.g.][]{Belokurov, Helmi, Naidu2020}. Consequently, the GSE has been studied qualitatively via analogs in cosmological simulations \citep{Bignone, Elias}, MW zoom-ins \citep{Fattahi, Grand}, and existing merger simulations \citep{Helmi, Koppelman}. These studies have achieved success in demonstrating how this merger produces the eccentric debris and reshapes the early Milky Way disk. 
   
   Building upon these findings, \citet{Naidu2022} made a tailored model for this merger by producing a grid of 500 idealized galaxy merger simulations with the GADGET code \citep{Springel05, Springel21} to identify a fiducial model that best matches the H3 survey data \citep{Conroy}. They produced a configuration to replicate the H3 data and explain disparate phenomena across the Galaxy. Recent evidence has reported that the GSE is likely a significantly gas-rich merger. This was highlighted by \citet{Ciuca} in their analysis of the age-metallicity relationship from the APOGEE-2 DR17 survey \citep{Abdurrouf}. They observed a notable hallmark of a gas-rich merger: an increase in [Mg/Fe] corresponding with a decrease in [Fe/H] around $\tau\approx12\,\text{Gyrs}$, closely aligning with GSE infall period. However, despite the success of the model proposed by \citet{Naidu2022}, it did not fully account for the considerable amount of gas introduced into the Galaxy by the GSE merger, which significantly contributes to the growth of both the high-$\alpha$ and low-$\alpha$ disks. 
   
   Based on measurements of precession and analyses of differences in other warp characteristics using mono-age samples, some studies have proposed an externally excited formation mechanism for the disk warp \citep{Poggio, Cheng}. However, these studies generally prefer a recent interactions with satellite galaxies that drive transient warps. \citet{Bosma} found that at least half of spiral galaxies exhibit warps, suggesting a long-lived and universal warp mechanism. A hallmark feature of $\Lambda$CDM cosmology is hierarchical assembly \citep[e.g.][]{White}, which means the warp might be formed in a universal way by galaxies merger.
    
   In this study, we advance our understanding of the Galactic warp by creating a gas-rich GSE merger simulation to reconstruct the morphology of our galaxy's disk.  In Section~\ref{sec:Model and simulation}, we provide a description of the simulation model and the initial conditions. In Section~\ref{sec:A long-lived warp}, we illustrate the long-lived and nonsteady warp model's evolution over time, comparing it with observational data at present. In Section~\ref{sec:Kinematic warp model}, we calculate the warp precession rate and demonstrate its lopsided feature. In Section~\ref{sec:Tilted DM halo}, we show that the disk is embedded in a live DM halo that is tilted and retrograde, possibly contributing to sustain the long-lived warp. Finally, we discuss and summarize our results in Section~\ref{sec:Discussion} and Section~\ref{conclusion}.

\section{Model and simulation} \label{sec:Model and simulation}

   \begin{table}
    \caption{Initial Conditions at $z\sim2$}\label{ICs}
    $$
      \centering    
         \begin{tabular}{llll}
            \toprule
            \noalign{\smallskip}
            Progenitor Parameters & Milky Way & GSE  & Units\\
            \noalign{\smallskip}
            \hline
            \noalign{\smallskip}
            Dark Matter mass  &  4.6   &  1.975  &  $\mathrm{1\times 10^{11}\,M_{\odot}}$ \\
            Dark Matter scale radius               &  26.7  &  19.1  &  $\text{kpc}$\\
            Stellar disk mass  &  60    &  5   &  $\mathrm{1\times 10^{8}\,M_{\odot}}$\\
            Stellar scale length            &  2     &  1.5   &  $\text{kpc}$\\
            Stellar scale height            &  1     &  0.75  &  $\text{kpc}$\\
            Gas disk mass    &  20    &  2  &  $\mathrm{1\times10^{9}\,M_{\odot}}$\\
            Gas scale length                &  6     &  6  & $\text{kpc}$\\
            Gas scale height                &  3     &  3  & $\text{kpc}$\\
            Bulge mass      &  1.4   &  N/A  & $\mathrm{1\times10^{10}\,M_{\odot}}$\\
            Bulge scale                     &  1.5   &  N/A  & $\text{kpc}$\\
            \noalign{\smallskip}
            \hline
            \noalign{\smallskip}
            Orbital Parameters\\
            \noalign{\smallskip}
            \hline
            \noalign{\smallskip}
            Eccentricity ($e$)                      &  0.75\\
            Initial distance                     &  141.2  & &$\text{kpc}$\\
            Normal vector of orbit ($\theta,\varphi$)                   &  (0, 125) & &$\text{degree}$\\
            MW and GSE spin $(\theta_{1},\,\theta_{2},\,\kappa)$                                 &  (125, 55, 180) & & $\text{degree}$\\
            Relative velocity                      & 73.02 &&$\text{km/s}$\\  
            \noalign{\smallskip}
            \hline
         \end{tabular}
    $$
   \end{table}


   In \citet{Naidu2022}, they made three different models to bracket mass range of $\mathrm{(2-7)\times 10^{8}\,M_{\odot}}$ for the GSE progenitor. Each model considered three scale lengths to account for the significant scatter in size at fixed mass observed at $z\sim2$. After simulation tests, they selected the fiducial model, determining the GSE progenitor mass to be $\mathrm{5\times 10^{8}\,M_{\odot}}$ and the scale length to be $\mathrm{1.5\times\text{SMR (size–mass relation)}}$. The DM halo mass was derived from the $z=2$ stellar mass-halo mass model in \citet{Behroozi}. The Milky Way was modeled as a combination of a thick disk and bulge with a total stellar mass of $\mathrm{2\times 10^{10}\,M_{\odot}}$, following the scale lengths in \citet{BHG}, and the DM halo was set to half the $z=0$ mass. More details of how these parameters are elaborated could be found in Section 4 in \citet{Naidu2022}. In summary, they created faithful representations of both the GSE and the Milky Way progenitors as follows: the GSE comprises a stellar disk and dark matter halo, while the Milky Way comprises a bulge, stellar disk, and dark matter halo.

   The numerical initial conditions (ICs) are generated from the disk initial conditions environment (DICE) code \citep{Perret14, Perret16}. The assumed density profiles for DM/stars/gas are input into DICE as distribution functions, which then generate Lagrangian particles with the Metropolis-Hastings Monte Carlo Markov Chain algorithm \citep{Metropolis}. In our simulation, We add a gas disk to each progenitors. The Milky Way progenitor comprises a bulge, a stellar disk, a gas disk, and a DM halo; where as the GSE progenitor is modeled with a stellar disk, a gas disk, and a DM halo. Both disk models follow an exponential + sech-z profile. The bulge is modeled using an Einasto profile \citep{Einasto} to replace the Hernquist profile \citep{Hernquist} used in \citet{Naidu2022}. The Einasto profile, which is similar to the Sérsic profile \citep{Sersic} but used for 3-D mass density \citep{Coe}, is a reasonable choice since the Sérsic profile is widely used for fitting the surface density of galaxies. The DM halo is modeled using the Hernquist profile.
   
    In Table~\ref{ICs}, we present the initial conditions of our simulation, where the virial masses of the Milky Way and GSE progenitors at $z\sim2$ are set to $\mathrm{5\times 10^{11}\,M_{\odot}}$ and $\mathrm{2\times 10^{11}\,M_{\odot}}$, respectively. Considering the higher gas fractions in galaxies at $z\sim2$, we set the GSE's gas fraction to 0.8 and the Milky Way's to 0.5, slightly higher than the typical values observed at $z\sim1.5$ \citep{Rodrigues}. We set the gas disk scale length for the Milky Way and GSE progenitors to $3R_{s}$ and $4R_{s}$ ($R_{s}$ is the scale radius of the stellar disk), respectively. These choices follow model assumptions made in \citet{Cox} and \citet{Rocha} that were used in simulation studies of galaxy mergers and metallicity gradients. The resolution mass for baryon particles is $\mathrm{1\times 10^{4}\,M_{\odot}}$ and for DM particles is $\mathrm{1\times 10^{5}\,M_{\odot}}$, and the initial relative velocity is calculated base on Keplerian orbit. The modeling approach utilized in this study is based on GIZMO \citep{Hopkins15}, in which we have implemented the module of star formation and feedback processes as in \citet{Wangj}, as star formation can convert gas particles into stellar particles and stars can turn back into gas via feedback. It is a variation of the smooth particle hydrodynamics code Gadget-3 and uses Adaptive Gravity Softening (more details in \citet{Hopkins18}). 

   As these structural choices play a secondary role to the total mass and orbital parameters described in \citet{Naidu2022}, they modeled the orbit with circularity: $\eta=0.5$, inclination: $\theta=15^{\circ}$. In \citet{Belokurov23}, they conducted a similar simulation broadly following \citet{Naidu2022}, but changed the inclination to $\theta=30^{\circ}$. Based on the fiducial model of \citet{Naidu2022}, we aimed to reconstruct both the GSE and the warp structure in our the simulation, leading to some differences in our orbital settings. Two galaxies with opposite disk spins were positioned in a a radially-biased and retrograde orbit. The initial distance between them was set equal to the sum of their virial radii. The eccentricity was set to 0.75, which is more radial than the moderate $\eta=0.5$ orbit. The polar ($\theta$) and azimuthal ($\varphi$) angles of the orbital plane's normal vector were set at ($0^{\circ},\,125^{\circ}$). This configuration indicates that GSE is inclined at $35^{\circ}$ relative to the MW disk plane in Z direction, slightly larger than \citet{Belokurov23}, and the spin angle is set to $(\theta_{1},\,\theta_{2},\,\kappa)=(125^{\circ},\,55^{\circ},\,180^{\circ})$, $\theta_{1}$ is the angle between the spin vector of the first galaxy and the orbital plane, $\theta_{2}$ is the angle between the spin vector of the second galaxy and the orbital plane, and $\kappa$ is the angle between the spin vector of the first galaxy and the second one (see Fig.2 in \citet{Perret14} for more details). After the gas-rich merger happened, gas particles could inherit its angular momentum from the orbital momentum of the merger and then be redistributed into a thin disk \citep{Barnes}. It will take several Gyrs for subsequent virialization phase to rebuild the disk \citep{Hopkins09, Hammer}, during which instabilities may develop some oscillations that we would like to search for. A more detailed summary of the constructed the GSE properties is provided in Appendix~\ref{sec:appendix}.

\section{A long-lived and nonsteady warp} \label{sec:A long-lived warp}

   \begin{figure*}
   \label{F1}
   \centering
   \includegraphics[scale = 0.2]{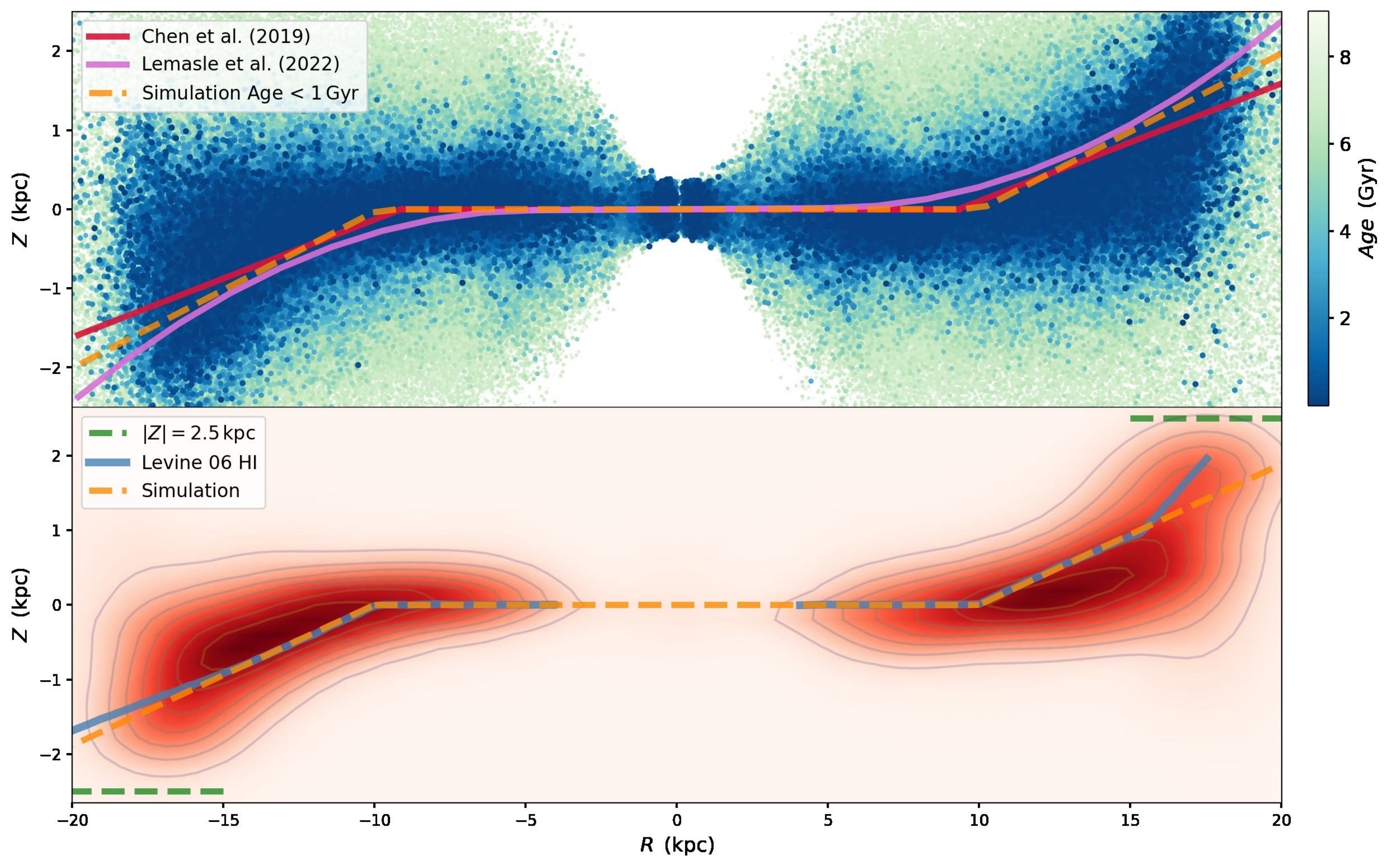}
   \caption{Distribution of simulated stars and gas in Galactocentric cylindrical coordinates at the present time (9.05 Gyrs since the simulation started). Negative/positive $R$ indicates the southern/northern warp. The top panel depicts the warp of the stellar disk, with a color bar indicating stellar ages. The bottom panel shows the gas warp in density projection map with contour profile. Here, the green lines represent a height of $|Z|=2.5\,\text{kpc}$, indicating the northern warp is stronger than the southern part. Comparative data are sourced from \citet{Chen}, \citet{Lemasle} and \citet{Levine}.}
              \label{F1}%
    \end{figure*}

   \begin{figure*}
   \label{F2}
   \centering
      \includegraphics[scale = 0.45]{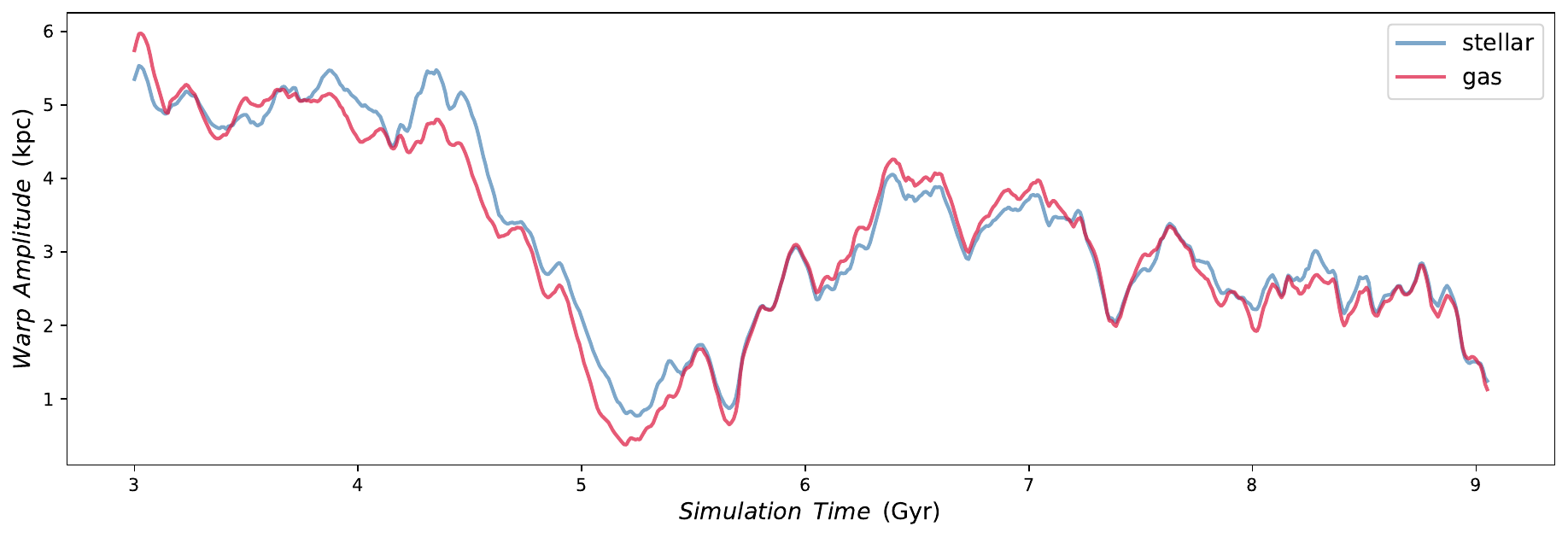}
      \caption{Evolution of stellar/gas disk warp amplitude as measured at $R=16\,\text{kpc}$, from 3 Gyr to 9.05 Gyr (as the merger occurred within the first 3 Gyrs), and the stellar disk amplitude is also fitting with stars younger than 1 Gyr. Both disks exhibit similar evolution. Notably, since the disk had not fully shaped before 3.25 Gyr, there are two approximate amplitude peak values observed around 4.3 Gyr and 6.4 Gyr.  
              }
         \label{F2}
   \end{figure*}
    
   The merger occurs rapidly, taking approximately 1 Gyr between the first and final pericenter passage, and is completed within the first 3 Gyrs, which is shown in the orbital decay profile panel of Figure~\ref{FigVibStab1}. As the GSE merger is estimated to have occurred approximately about 8-11 Gyrs ago \citep{Belokurov, Helmi}. In Figure~\ref{F1}, we provide an edge-on perspective of the galactic disk, selecting particles based on their circularity: $\epsilon=L_{Z}/L_{Z, max(E)}>0.65$. This selection criterion is applied at 9.05 Gyrs after simulation began, corresponding to what we expected as the present ($z\sim0$). This time aligns with Auriga 18 (Au18), a Milky Way-like disk galaxy from the Auriga project \citep{Grand17}, which is frequently referenced in GSE research. This visualization distinctly reveals an S-shaped warp in both the stellar and gas disks, consistent with observational data. Stars are color-coded based on their ages; elder stars are represented with lighter green markers, while younger stars are depicted with more vibrant blue markers. The warp is the most pronounced for the youngest stars and spans multiple age groups \citep{Chen, Cheng, Huang}. The geometric shape of the warp can be approximated by a power-law warp mode:
      \begin{equation}
      \label{f1}
      Z_{w}(R\ge R_{w}) =a(R-R_{w})^{b} sin(\phi -\phi_{w}).\\
      \end{equation}
      \begin{equation}
      \label{f2}
      Z_{w}(R< R_{w}) =0.\\
      \end{equation}
   
   Here, $R$ and $Z_{w}$ are Galactocentric cylindrical coordinates, $a$ represents the amplitude of the warp, $b$ is the power-law index, $\phi_{w}$ is the polar angel of warp's lines of node (LON hereafter) indicates the orientation of the warp and $R_{w}$ is the onset radius of the warp. This model was employed by \citet{Chen} to accurately fit the Cepheid tracers, prompting us to select stars within the young stars' age range ($age<1 Gyr$) in the simulation for comparison with the Cepheid data. As \citet{Chen} notes, the parameters of $a$, $b$ and $R_{w}$ show a clear correlation, whereas $\phi_{w}$ is a more independent parameter in the geometric warp model. Consequently, while different studies might yield varying $a$, $b$ and $R_{w}$ values, $\phi_{w}$ tends to be more consistent across measurements of the Galactic warp. In our analysis, we refrain from comparing $\phi_{w}$ with observational data, as its value can change based on the reference frame specified in the simulation, so we define the $\phi_{w}$ as the y-axis making the orientation similar to that of the MW. 
   
   In the upper panel of Figure~\ref{F1}, we fitted the warp model in simulations and found that the linear model ($b=1$) closely matches the observational data, the simulation's maximum warp amplitude is marked by a yellow dashed line, aligning closely with the Cepheid data \citep{Chen, Lemasle}. The gas density projection map with contour profile is displayed in the lower panel, in central $\sim1$ kpc where the formation of the Galactic bar creates a noticeable gap in the inner region. Both sides of the gas warp align closely with the H\uppercase\expandafter{\romannumeral1} data from \citet{Levine} within a 15 kpc range. Beyond this region, observational data reveal that the northern section (with positive peak) exhibits a higher amplitude, whereas the southern part (with negative peak) displays a lower amplitude. Nonetheless, the solid line remains within high gas density region. Despite these minor discrepancies, the overall results are deemed satisfactory. Previous research utilizing different tracer types has noted different observational systematics in the measured amplitudes \citep{He}.

   Previous studies have reported the asymmetry of the warp in both gas and stellar components, with findings suggesting that the northern warp is more pronounced than its southern counterpart  \citep{Levine, Reyle, Amores}. Accordingly, we plot horizontal lines to represent the height of $|Z|=2.5\,\text{kpc}$ in the bottom panel of Figure~\ref{F1}. This visualization shows the warp's asymmetry in our simulation: the southern part falls slightly below this height, while the northern part is consistent with observational data. It should be noticed that in our simulation, the disk warp evolves continuously, such asymmetry could occasionally disappear or suddenly intensify.
          
   Figure~\ref{F2} illustrates the time evolution of the stellar/gas disks warp's amplitude, calculated by fitting Eq.~\ref{f1} to the warp every 0.01 Gyr since $t=3\,\text{Gyr}$ that after the merger has been completed and identifying the maximum amplitude at $R=16\,\text{kpc}$, a radius near the edge of the disk in our simulation, and the stellar disk is fitting with stars younger than 1 Gyr. Both disks exhibit similar evolution in Figure~\ref{F2}, as gas particles can convert into young stars in the simulation, such evolution trend highlights the importance of gas component in reconstructing the Galactic warp. The thin disk was gradually formed after the merger in our simulation, corresponding to a look-back time of about 6 Gyr, which is close to the observational time frame when the thin disk gradually formed about 7 Gyr ago \citep{Snaith}, disk had not fully shaped before 3.25 Gyr, which leading to chaotic conditions and potentially inaccurate measurements prior to this point. Afterwards, the warp reached its initial peak at approximately 4.3 Gyr, decreased until around 5.2 Gyr, saw another peak at 6.4 Gyr, and then generally diminished to present. Thus, we can identify roughly many evident extremum values in Figure~\ref{F2}, indicating that the warp is a long-lived and nonsteady structure.

\section{Kinematic warp model} \label{sec:Kinematic warp model}

   \begin{figure*}
   \label{F3}
   \centering
      \includegraphics[scale = 0.45]{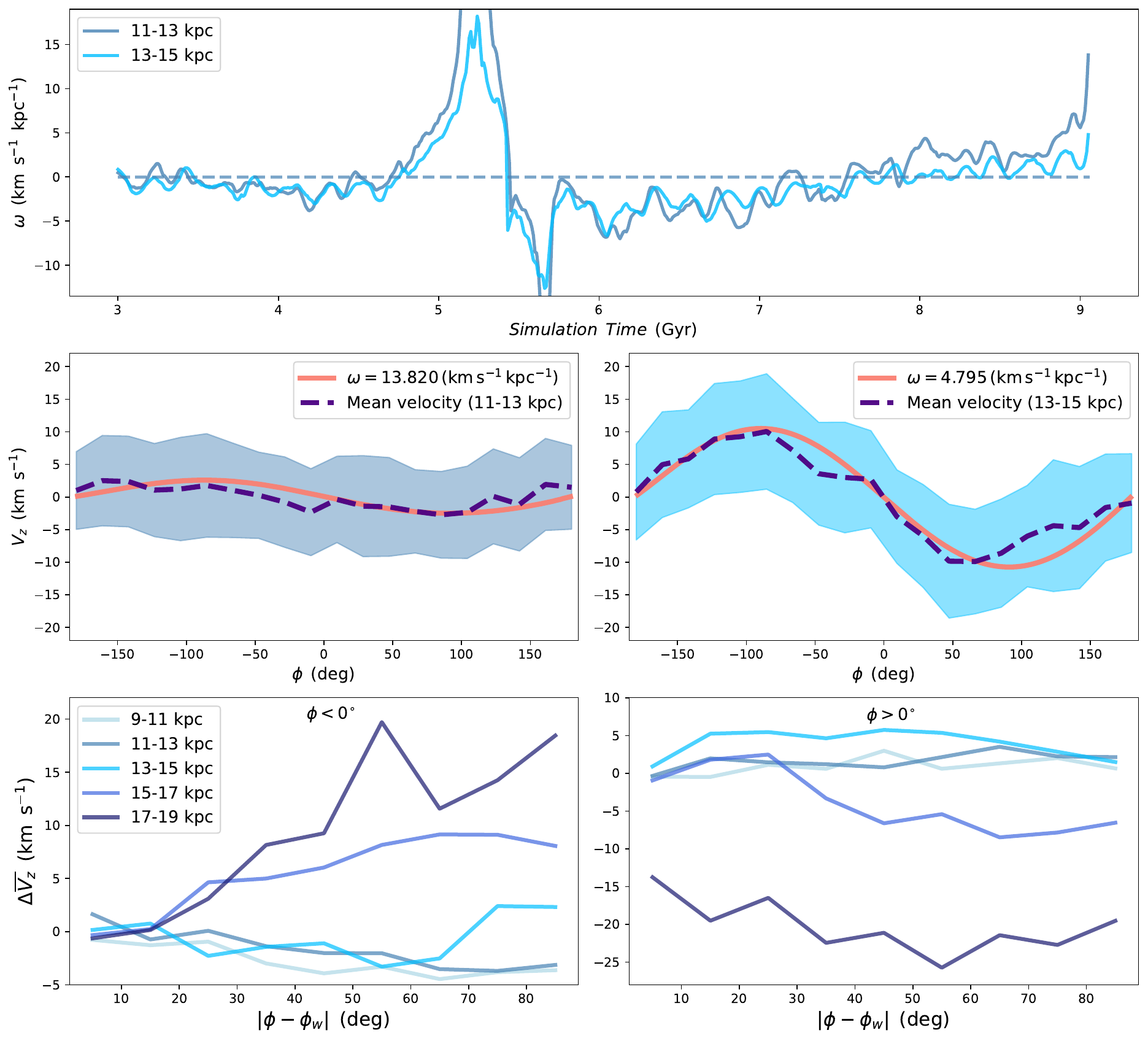}
      \caption{$Top\enspace panel$: The evolution of warp precession as measured with young stars within the radial range of $R=11\sim13\,\text{kpc}$ and $R=13\sim15\,\text{kpc}$ from 3 Gyr to 9.05 Gyr. $Middle\enspace panel$: The transparent band shows the distribution trend of the young stars' velocity in Z direction versus azimuth within the region of $R=11\sim13\,\text{kpc}$ (left) and $R=13\sim15\,\text{kpc}$ (right), the purple line is the mean velocity, and the pink  line indicates the fitting precession with kinematic warp model. $Bottom\enspace panel$: The vertical velocity differences of stars on both sides of $\phi_{w}$ vs. azimuth angle separation from $\phi_{w}$. The azimuth angle are binned in $10^{\circ}$, and each radial annulus are plotted with different colors.  
              }
         \label{F3}
   \end{figure*}

   In section~\ref{sec:A long-lived warp}, we discuss the evolution of the Galactic warp, highlighting how its rapid amplitude changes underscore the dynamical nature of warps. This dynamism is crucial for unlocking insights into the formation history of galaxies and the mass distribution of their halos. The changing geometry of the warp can be succinctly described by the variation in the direction of its LONs at a precession rate $\omega$. This is expressed as $\phi_{w}(t) = \phi_{w,0}+\omega t$, where $\phi_{w,0}$ represents the current position of the LONs. A time dependent model of warp is provided by \citet{Poggio}, described as:
      \begin{equation}
      \label{f3}
      \overline{V_{Z}}(R,\phi,t=0)=(\frac{\overline{V_{\phi}}}{R} -\omega(R))h_{w}(R)cos(\phi-\phi_{\omega})+\frac{\partial h_{w}}{\partial t}sin(\phi-\phi_{w}).\\
      \end{equation}
      
   Where $h_{w}(R)=a(R-R_{w})^{b}$, as described in Eq.~\ref{f1}, $\overline{V_{Z}}$ represents mean vertical velocity and $\overline{V_{\phi}}$ is mean azimuthal velocity. To simplify the calculations, we neglect the time derivative term $\frac{\partial h_{w}}{\partial t}$ and assume $\omega$ does not vary with radius. Although the warp is evolving, our calculations indicate that this approximation does not significantly affect the precession results. With this assumption, the model could be described the same as Equation 7 in \citet{Poggio}:
      \begin{equation}
      \label{f4}
      \overline{V_{Z}}(R,\phi)=(\frac{\overline{V_{\phi}}}{R} -\omega)h_{w}(R)cos(\phi-\phi_{w}).\\
      \end{equation}
      
   With this kinematic model, after determining the geometric shape through fitting with Eq.~\ref{f1}, we can plug in the parameters of $h_{w}$ and $\phi$ into Eq.~\ref{f4} and measure the procession rate $\omega$ using kinematic data from the simulation. Here we select the young stars ($age<1Gyr$) within the radial range of $R=11\sim13\,\text{kpc}$ and $R=13\sim15\,\text{kpc}$, plotting the evolution of $\omega$ over time in the top panel of Figure~\ref{F3}. The dashed line represents the non-precession state, while positive and negative values of $\omega$ indicate prograde and retrograde warp precession relative to disk rotation, respectively. The trend of $\omega$ fluctuates between positive and negative values, for example the $R=13\sim15\,\text{kpc}$ precession trend ranges approximately from -13 to 18 $\mathrm{km\,s^{-1}\,kpc^{-1}}$. Although there are periods where the precession rate is notably high in either the prograde or retrograde direction, it typically remains within a range of $-5\sim5\mathrm{km\,s^{-1}\,kpc^{-1}}$. A significant shift in the precession rate is observed around $t=5.5\,\text{Gyr}$ in Figure~\ref{F2}, which can be attributed to the exceptionally low warp amplitude during this period. Comparing both the evolution trend of precession and amplitude in Figure~\ref{F2}, we could find that both two consistently reach their extremum simultaneously, indicating a synchronous evolution. In the middle panel of Figure~\ref{F3}, the transparent band illustrates the velocity trend of simulated stars younger than 1 Gyr within the radial ranges of $R=11\sim13\,\text{kpc}$ (left) and $R=13\sim15\,\text{kpc}$ (right) at the present time, with the dashed line representing the mean velocity. This clearly demonstrates the warp's dynamical characteristics. In our simulation, it shows a high prograde precession in the range of $R=11\sim13\,\text{kpc}$. Incorporating all geometric model parameters into Eq.~\ref{f4}, we obtain a precession rate of 13.820 and 4.795 $\mathrm{km\,s^{-1}\,kpc^{-1}}$ for two ranges, this value is close to some current precession observational data. It can be found that the fitting curve does not perfectly cover the mean velocity. This discrepancy arises because the vertical velocity is asymmetric about the longitude of peak vertical velocity, indicating that the warp is lopsided \citep{Amores, Cheng, Li23}. In the bottom panel of Figure~\ref{F3}, refer to \citet{Cheng} we measure the velocity asymmetry by subtracting the median vertical velocity of stars on one side of $\phi_{w}$ from its complement on the other side at the same azimuthal separation in each radial annulus. A gradually increasing difference in vertical velocity is found when the azimuth angle moving away from the velocity peak, consistent with the feature reported in \citet{Cheng}. This trend may become more pronounced in the outer disk, as the disk is asymmetric we can also found the lopsided feature is more evident in the $\phi<0^{\circ}$ area.
   
   In \citet{Poggio}, precession was calculated using 12 million red giant stars from Gaia DR2 \citep{Gaia}, yielding a precession rate of $\mathrm{10.86\pm 0.03_{stat}\pm3.20_{syst}\,km\,s^{-1}\,kpc^{-1}}$ in the direction of Galactic rotation. Recently, the similar model was employed in \citet{Zhou}, using 134 Cepheids and combined with the line-of-sight velocity (RV) from Gaia DR3 \citep{Gaia3}, determining a lower warp precession rate of $\mathrm{4.9\pm1.6\,km\,s^{-1}\,kpc^{-1}}$, our simulation is very close to these observation in different ranges. Another study by \citet{Chr} also suggested a low precession rate and cannot exclude a non-precessing warp of MW. Consequently, determining precession accurately is hindered by the lack of comprehensive spatial and kinematic data, compounded by the need for reliable tracers. In Figure~\ref{F4}, we plot the variation of the precession rate with Galactocentric radius R. As shown, our simulation results align well with other studies in the literature \citep{Poggio, Zhou, CC} for $R<\sim15\,\text{kpc}$. Both \citet{Dehnen23} and \citet{CC} reported a prograde precession that decreases with increasing Galactocentric radius, a trend that our simulation also captures. In our results, the precession rate continues to decrease, eventually turning negative ($R\sim15.5 \,\text{kpc}$), indicating a transition to retrograde precession. As the outer disk is more susceptible to be disturbed, which is also likely to be influenced by Sgr. and LMC \citep{Stelea}, which might make precession rate rise in observation beyond 15 \text{kpc}.

   \begin{figure*}
   \label{F4}
   \centering
      \includegraphics[scale = 0.5]{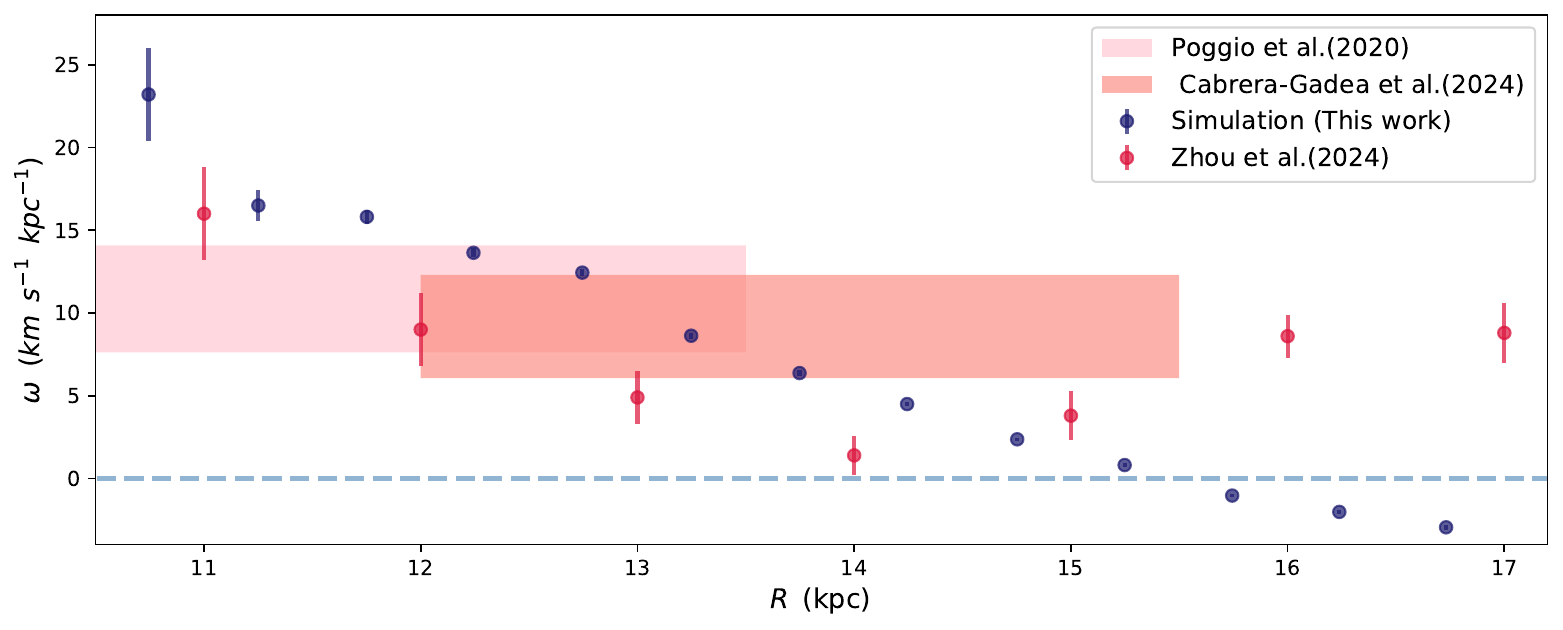}
      \caption{Comparison between the obtained values of warp precession rate varies with Galactocentric radius R in our simulation (purple circle), together with other values from the literature. 
              }
         \label{F4}
   \end{figure*}

   Theoretical and numerical studies of misaligned non-spherical halo potentials imply that the direction of precession can be influenced by the DM halo shape, as the prolate or prolate-like halo can rise prograde precession, while oblate-like halo should produce a retrograde warp precession \citet{Poggio}. However, warps in oblate halos are not appropriate to explain the observed warps or they could not sustain a prominent warp \citep{Dubinski, Jeon, Ideta}. Typical precession periods from such models or those produced by a misaligned outer torus of later accreted material \citep{Shen, JB, Jeon}, are approximately between 4 Gyr and 40 Gyr, corresponding to warp precession rates between $\mathrm{1.5\,km\,s^{-1}\,kpc^{-1}}$ and $\mathrm{0.1\,km\,s^{-1}\,kpc^{-1}}$. In Figure~\ref{F3}, the precession rate determined after the merger significantly surpasses all predicted values throughout the simulation period. Since GSE merger led to a oblate halo which could not account for the observed characteristics of warp, and a percession value greater than $\mathrm{10\,km\,s^{-1}\,kpc^{-1}}$ indicates a likely transient response observed in the outer disk. Such a response could be attributed to interactions with a satellite galaxy \citep{WB, Laporte}, a scenario not included in our simulation. Given the warp's persistence over an extended period and its exhibition of both high prograde and retrograde precession post-merger, unraveling the underlying dynamical mechanisms poses a challenge.

\section{Tilted and Retrograde DM halo} \label{sec:Tilted DM halo}

   \begin{figure*}
   \label{F5}
   \centering
      \includegraphics[scale = 0.45]{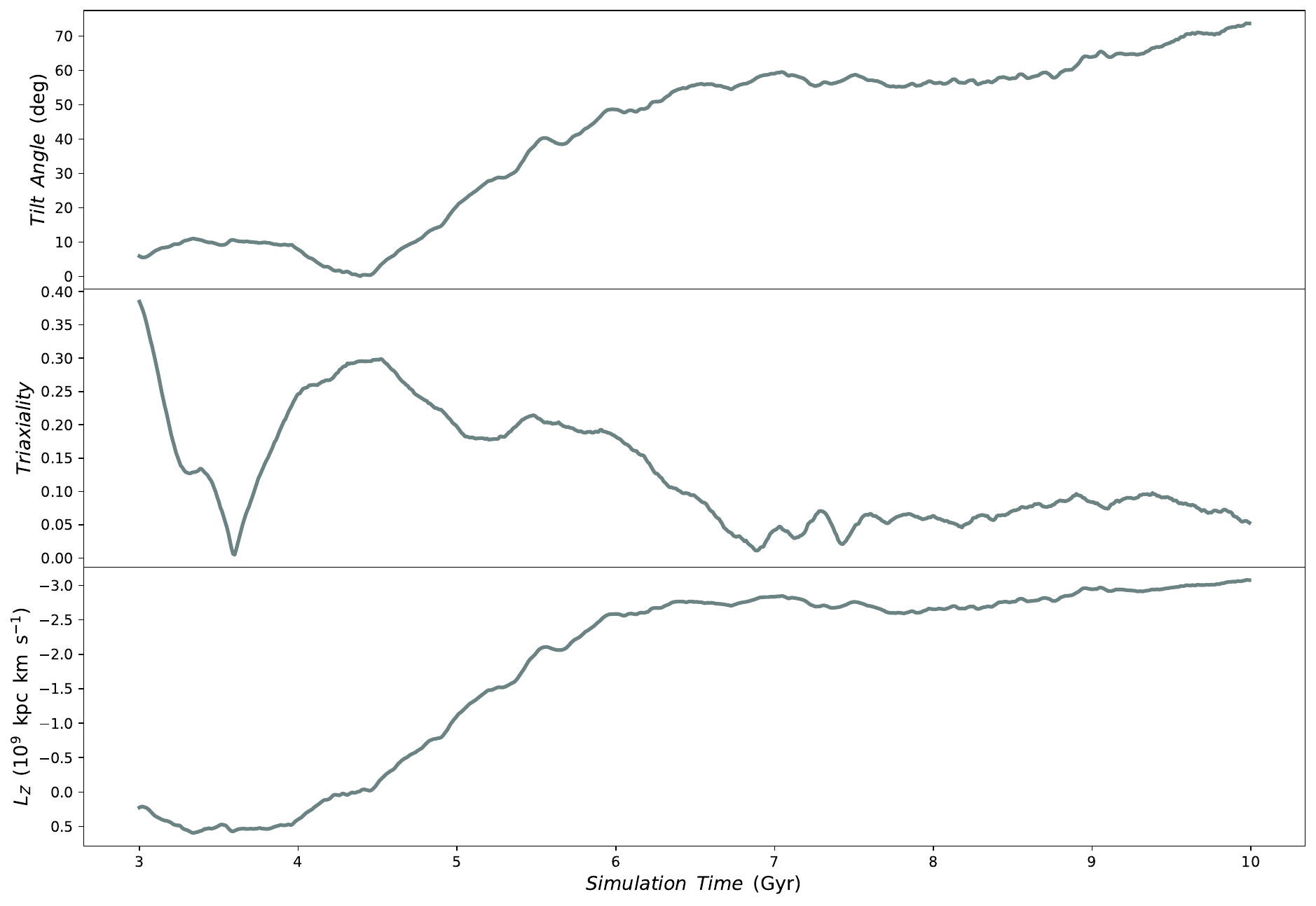}
      \caption{$Top\enspace panel$: Illustrating the evolution of the DM halo's tilt angle with respect to the disk. $Middle\enspace panel$: DM triaxiality variations with time, with values predominantly less than 0.3, indicating an oblate shape. $Bottom\enspace panel$: Net angular momentum about the Z-axis $L_{Z}$ of the DM halo as a function of time. Negative value indicates a retrograde halo.
              }
         \label{F5}
   \end{figure*}

   \begin{figure*}
   \label{F6}
   \centering
      \includegraphics[scale = 0.45]{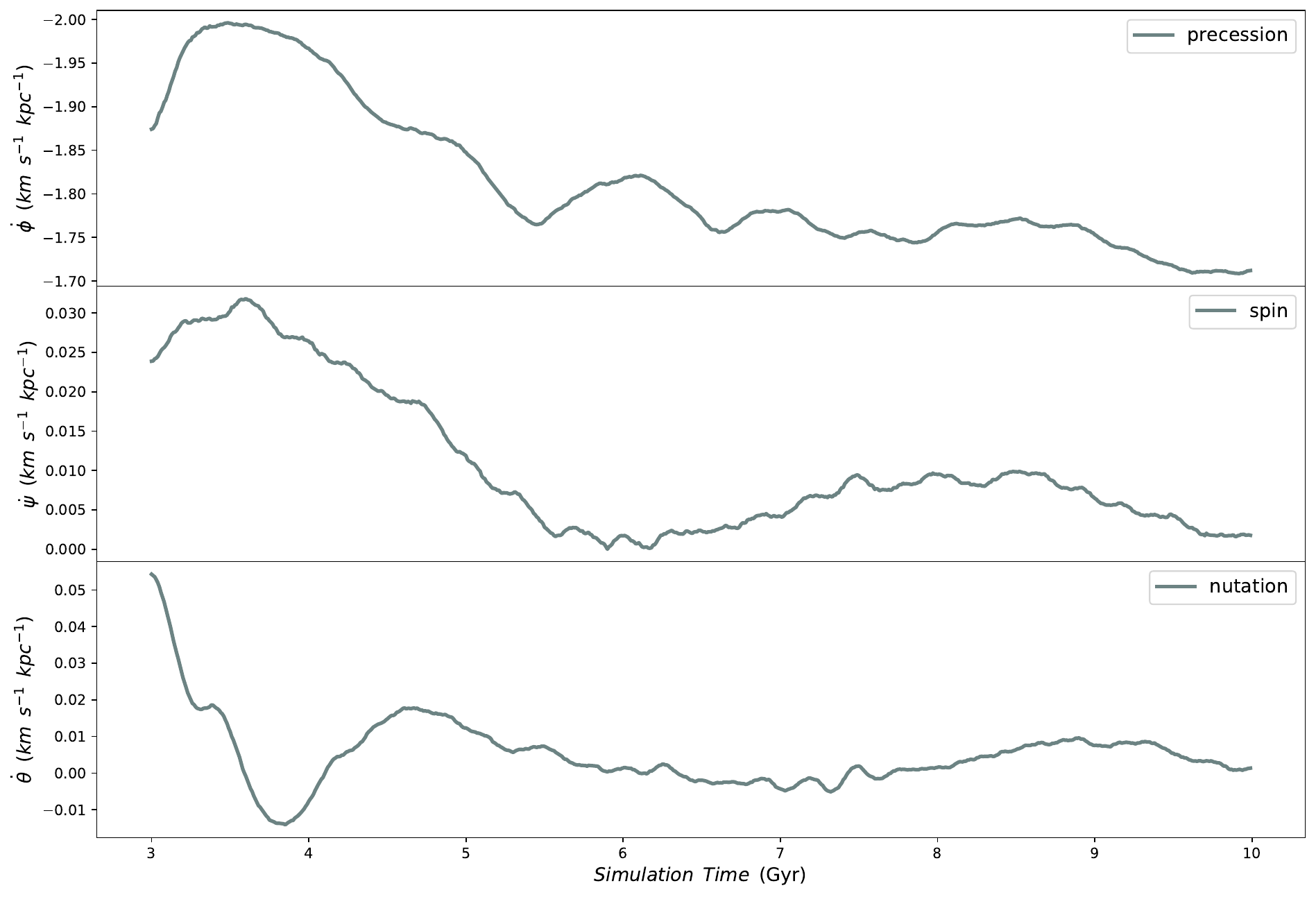}
      \caption{Three velocities vary with time in different panels. The precession rate is retrograde relative to the disk. The spin velocity is along the body-fixed z axis. The nutation velocity is along the {x}' axis, X axis after the first rotating with the $\phi$ angle.  
              }
         \label{F6}
   \end{figure*}

    \citet{Han22} discovered that the stellar halo, spanning a Galactocentric radius of 5-50 kpc, is tilted approximately $\sim25^{\circ}$ with respect to the disk plane. Following this discovery, they executed a numerical simulation utilizing a gravitational potential model of a galaxy. In their approach, 30$\%$ the dark halo mass adopts a triaxial distribution, which is similarly tilted by $25^{\circ}$ above the Galactic plane towards the Sun. Their results suggest that the warp and the flare of the Galactic disk might be related to the tilted halo \citep{Han23}.

    If so, our simulated disk might also be embedded in a tilted DM halo. We select the DM particles within $10\,\text{kpc}<r<50\,\text{kpc}$ of Galactocentric radius, with the mass and position of each particles we can calculate the moment of inertia tensor. Solving for the eigenvector-eigenvalue pairs of inertia tensor, we can find the three principal axes of rotation and their respective moments of inertia. The major axis has the minimum moment of inertia and minor axis has the maximum moment of inertia. The length of the principal axes $r_{i}$, $i\in \left \{ a,b,c\right \}$ has the relation with the moments of inertia $I_{i}$ as following \citep{Peraire}:
    \begin{equation}
      \label{f5}
      r_{i}^{2}\propto \frac{-I_{i}+I_{j}+I_{k}}{2}.\\
      \end{equation}
    Once we obtain the moments of inertia, we can calculate the major-to-intermediate and major-to-minor axes ratios as follows:
    \begin{equation}
      \label{f6}
     \frac{r_{b}}{r_{a}} :\frac{r_{c}}{r_{a}} =\sqrt{\frac{-I_{b}+I_{a}+I_{c}}{-I_{a}+I_{b}+I_{c}}}: \sqrt{\frac{-I_{c}+I_{a}+I_{b}}{-I_{a}+I_{b}+I_{c}}}.
     \end{equation}  
    
    Then we can calculate the triaxiality with $T=(1-p^{2})/(1-q^{2})$, where $p$ and $q$ represent the ratios of major-to-intermediate and major-to-minor axes, respectively. $T>0.6$ denotes a prolate halo, and $T<0.3$ indicates an oblate halo. By measuring the orientation of the axes (with the disk plane's normal vector aligned parallel to the Z-axis), we are able to determine the tilt angle. Our results indicate that the major and intermediate axes are degenerate, a scenario also observed in \citet{Kazantzidis} and \citet{Shao}. This degeneracy renders the measurement unstable if these axes are chosen. However, the minor axis is nondegnerate and can be reliably used to measure the tilt angle of the DM halo with respect to the disk plane.

    In the middle and bottom panels of Figure~\ref{F5}, we illustrate the triaxiality with particles in $10\,\text{kpc}<r<50\,\text{kpc}$ and the net angular momentum of DM halo within 50 kpc. Following the GSE merger, the simulation halo appears an oblate profile, in alignment with \citet{Iorio}, and maintains a negative angular momentum, indicating a retrograde halo with respect to the disk. \citet{Joshi} suggest that the vertical displacement of the disk in a retrograde halo will be stronger than in a prograde halo, leading to a more organized warp. For a simple heuristic argument that  dynamical friction tries to bring the angular momentum vectors of the halo and disk into alignment, which means the warp will be strongly damped in co-rotating halo but enhenced in counter-rotating halo \citep{nt}. Thus, the retrograde halo resulting from GSE merger could sustain a long-lived warp. 
    
    The top panel of Figure~\ref{F5} illustrates the time evolution of the DM halo tilt angle with respect to the disk, showing a continuously increasing trend, and there is no evident correlation with the amplitude evolution. In the simulation of \citet{Han23}, they fixed the halo's tilt angle at $25^{\circ}$ and found that the warp need 1.5 Gyr to reach a steady-state amplitude after that there is no significant change. This suggests that a live halo could facilitate a fast and nonsteady evolution of the warp. The halo's tilt angle continues to increase following the merger, which, as suggested by \citet{Han23}, should result in an increasing warp amplitude. However, our simulation could produce warp with a low amplitude at $t = 5.2\,\text{Gyr}$ and show irregular evolution as illustrated in Figure~\ref{F2}. Therefore, we posit that in galaxy models with a more realistic cosmological setting, the actual mechanisms driving the warp evolution are likely to be more intricate than suggested by idealized simulations based solely on gravitational potential.

    Such changes in the tilted angle suggest that the DM halo might also exhibit rotational precession. We can approximate the DM halo as a rotating rigid body in the absence of external moments. In this scenario, we identify the body-fixed principal axes for the DM halo, resulting in a moment of inertia tensor in the simplified form of $[I]=diag(I_{xx},I_{yy},I_{zz})$. Consequently, the angular momentum vector can be expressed as  $\mathbf{H_{G}}=\left \{  0,0,H_{G} \right \}$. 
    
    To describe the precession of the DM halo, we select the particles within 50 kpc and establish a body-fixed coordinate system by applying Euler angles with the rotation order $R_{Z}R_{X}R_{Z}$ to translate the X-Y-Z absolute reference frame system which is initially defined in the simulation box. The free motions of the DM halo is then characterized by changes in these three Euler angles, which can be calculated as follows \citep{Peraire}:
    \begin{equation}
      \label{f7}
       \dot{\phi} = H_{G}(\frac{cos^{2}\psi}{I_{yy}}+ \frac{sin^{2}\psi}{I_{xx}}).
     \end{equation}
    \begin{equation}
      \label{f8}
       \dot{\theta} = H_{G}(\frac{1}{I_{xx}} - \frac{1}{I_{yy}})sin\theta sin\psi cos\psi.
     \end{equation}
    \begin{equation}
      \label{f9}
       \dot{\psi} = H_{G}(\frac{1}{I_{zz}} - \frac{cos^{2}\psi}{I_{yy}}-\frac{sin^{2}\psi}{I_{xx}} )cos\theta.
     \end{equation}

     The $\psi,\phi,\theta$ are described as ``roll", ``yaw" and ``pitch" angle, and $\dot{\psi},\dot{\phi},\dot{\theta}$ are spin, precess and nutate rate respectively. Under the condition that the DM halo exhibits an oblate profile, we align the minor axis with the z-axis in the body-fixed coordinate system, the evolution of three velocities are shown in Figure~\ref{F6}. The precession rate of DM halo is along the original Z axis. Given that the angular momentum of the dark matter (DM) halo is opposite to that of the disk, the DM halo undergoes retrograde precession relative to the disk. This interaction may cause the disk warp to also exhibit retrograde precession, as shown in Figure~\ref{F3}, where the precession remains predominantly retrograde throughout most of the period.

\section{Discussion}\label{sec:Discussion}

    It is crucial to note that in our simulation, one of the important parameters influencing warp creation is the azimuthal angle of the normal vector of the orbital plane. Specifically, we chose a slightly larger angle of $125^{\circ}$, which resulted in a particularly small part the GSE debris following a high-energy orbit. These stars do not align well with current observations, suggesting that the simulation configuration may not be the most accurate model for reconstructing the GSE. Additionally, our work does not consider other potential mechanisms for the creation of warp, such as interactions with satellite galaxies, or other  minor/mini infall events. Therefore, this suggests that single major merger model could not simultaneously fit the both the GSE and the disk warp very accurately. 

    In Section~\ref{sec:Kinematic warp model}, we calculated the precession rate versus time, revealing the precession can alternate between prograde and retrograde. In the present, it aligns well with observational data for $R<\sim15\,\text{kpc}$. However, beyond this range our simulation shows a transition to retrograde precession, potentially inconsistent with current observational data in this range. However, precise measurements necessitate extensive kinematic data. Additionally, gravitational interactions with a satellite galaxy, not accounted for in our simulation, could induce a prograde precession. Given the extensive evolutionary history of the Milky Way, it's plausible that other mechanisms affecting the precession rate have occurred. Therefore, incorporating additional models partway through this period may provide a more comprehensive understanding of these dynamics.

    In Section~\ref{sec:Tilted DM halo}, we find our disk is embedded in a time-evolving tilted and retrograde DM halo. In \citet{Han23}, their research focused exclusively on a fixed tilt angle of the DM halo, neglecting the long-term dynamical mechanisms that emerge from mergers. As they suggest that the warp requires 1.5 Gyrs to reach steady-state, our simulations indicate that such an extended period could result in significant changes within the DM halo. Therefore, we posit that variations of the tilt angle relative to the disk might influence the formation of warp. Notably, the time evolution of the DM halo tilt angle does not exhibit a clear correlation with the observed changes in warp amplitude. This implies that in a comprehensive galaxy model, the impact of a tilted DM halo on warp formation may not be as straightforward as predicted by idealized gravitational potential models. Consequently, the Galactic warp model appears to necessitate a series of controlled experiments for further investigation, which we are preparing to conduct in our next study.

\section{Conclusion}\label{conclusion}

     We conducted a hydrodynamical simulation of gas-rich GSE merger, positioned on reconstructing the GSE debris, successfully producing the Galactic warp amplitude and precession that aligns well with observational data, discovering that the warp is long-lived, nonsteady, asymmetric, lopside, exhibiting both prograde and retrograde precession rate after the completion of the merger. We found significant changes in the DM halo as a consequence of the major merger, which created a live, oblate, tilted, and retrograde DM halo. Specifically, the tilted and retrograde characteristics of the DM halo appears to sustain the warp.

\begin{acknowledgements}

   This work was supported by the National Natural Science Foundation of China (NSFC Nos.11973042 and 11973052). We thank Xiaodian Chen for sharing the data. We are grateful to Phil Hopkins and Jianling Wang who kindly shared with us the access to the Gizmo code. We are grateful for the support of the International Research Program Tianguan, which is an agreement between the CNRS in France, NAOC, IHEP, and the Yunnan Univ. in China .
\end{acknowledgements}

%



\appendix
\section{Characteristic of simulated GSE} \label{sec:appendix}
       \begin{figure*}
       \centering
          \includegraphics[scale = 0.125]{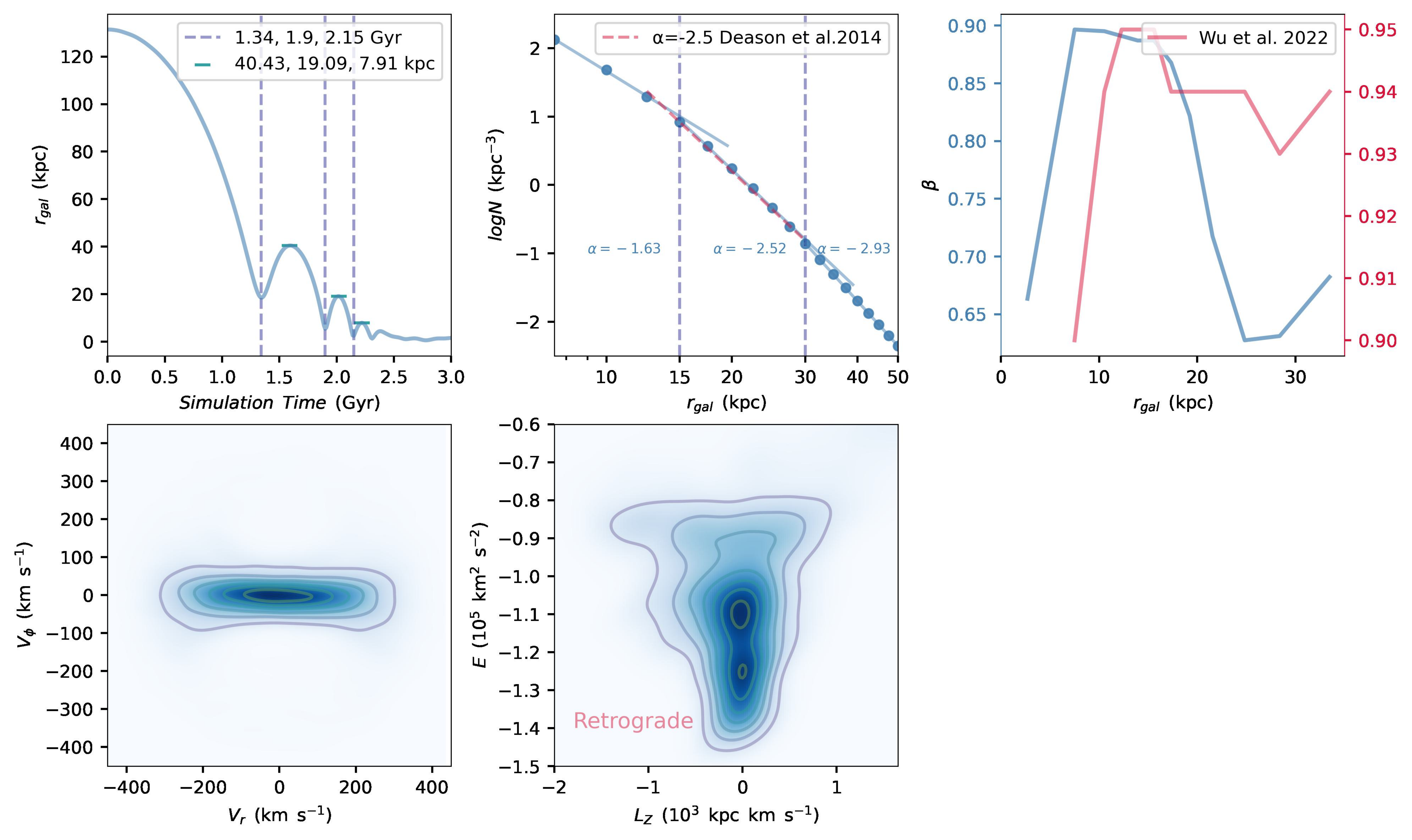}
          \caption{$Top\enspace left\enspace panel$ shows the orbital decay profile, the merger completed within the first 3 Gyrs. $Top\enspace middle\enspace panel$ shows all-sky density profile of GSE (blue points/lines), there are two breaks in our simulation and the red dash line is observational stellar halo profile \citep{Deason}. $Top\enspace right\enspace panel$ is the anisotropy parameter $\beta$ of the GSE debris with Galactocentric distance in a spherical coordinate system, comparing with observational data. $Bottom\enspace panels$ are the most representative characteristic of GSE potted with contour profile, the “sausage” structure is in the left panel and the distribution of the particle binding energy vs. angular momentum in the right panel. Here the negative value of $V_{\phi}$ and $L_{Z}$ means that the stars on a retrograde orbit. Except for the top left panel, all the other panels are plotted at 9.05 Gyr.
          }
             \label{FigVibStab1}
       \end{figure*}

    Here we list a summary of our GSE model in Figure~\ref{FigVibStab1}. While most of the galaxy model parameters are adapted from \citet{Naidu2022}, our model features a significantly different orbit configuration. Nevertheless, our model successfully reconstructs the global characteristics of the GSE debris. The top-left panel illustrates the orbital decay profile. Owing to the large eccentricity of the orbit setting, the merger occurs rapidly, taking approximately only 1 Gyr between the first and final pericenter passages. The top-middle panel shows the all-sky density profile of the GSE, the power-law coefficients are given by $\rho \propto r_{gal}^{\alpha}$ profile. As \citet{Naidu2022} proposed, we also adopted a double-break density profile which is used widely for the inner halo ($r_{gal}<30\,\text{kpc}$), and noticed that there are two breaks in $r_{gal}=15\,\text{kpc}$ and $r_{gal}=30\,\text{kpc}$ respectively. The break at $15\,\text{kpc}$ is a critical feature of the GSE merger \citep{Naidu2022} and it also roughly the radius where the disk warp reaches maximum amplitude, which may hint a correlation between the GSE and warp. Since the GSE is by far the most dominant component of the inner halo, we expect the overall halo density profile at this range could be fitted in our simulation, we find the slope of the GSE density proﬁle between two breaks is a good match to that found for the inner halo \citep{Deason}, and the third coefﬁcients after the second break fits well with that of \citet{Ye}, as they report that the second break at $27.18\,\text{kpc}$ with $\alpha=2.86$. Meanwhile, the top-right panel displays the anisotropy parameter $\beta$ of GSE debris with Galactocentric distance in a spherical coordinate system, which defined as:
      \begin{equation}
      \beta = 1-\frac{\sigma_{\theta}^{2}+\sigma_{\phi}^{2} }{2\sigma_{r}^{2}}. \\
      \end{equation}
    where $\beta$ quantiﬁes the degree of velocity anisotropy of a system of stellar orbits, where $\sigma_{i}$ are the velocity dispersions in spherical coordinates. The observational data is from \citet{Wu},  derived from the LAMOST \citep{Zhao06, Zhao12, Luo, Cui} K-giant Sgr-removed stellar halo. There exists a disparity between the parameter values in the data and the simulation results. However, both exhibit a consistent trend characterized by an increasing anisotropy parameter within $17\,\text{kpc}$ and a relatively stable to decreasing trend between approximately $17-28\,\text{kpc}$, and increasing after that. The bottom panels show most prominent features of GSE in both $V_{\phi}-V_{r}$ panel and the $E-L_{Z}$ panel. We can find the “sausage” structure in velocity panel which was first discovered in \citet{Belokurov} as the “Gaia-Sausage” came from, the contour profile tends to have negative values (indicating retrograde orbits in our setting) on both sides, implying a slight overall retrograde motion. The same feature also reflected in $E-L_{Z}$ panel, where the majority of stars are situated in the central region, and the profile similarly trends toward negative values.




\end{document}